\documentclass[prl,aps,showpacs,floatfix,floats,superscriptaddress,nobalancelastpage,twocolumn]{revtex4-1}
\usepackage{graphicx}
\usepackage{graphics}
\usepackage{latexsym,amsmath,amssymb,bm,euscript}
\usepackage{color}
\usepackage{dcolumn}
\usepackage{bm}
\usepackage{epstopdf}
\usepackage{times}
\usepackage{multirow}
\usepackage{wasysym}
\usepackage{subfigure}
\usepackage{bbm}
\def\ra{\rangle}
\def\la{\langle}

\def\Hc{{\rm H.c.}}

\begin{document}

\title{Short-ranged interaction effects on $Z_2$ topological phase transitions}
\author{Hsin-Hua Lai}
\affiliation{National High Magnetic Field Laboratory, Florida State University, Tallahassee, Florida 32310, USA}
\author{Hsiang-Hsuan Hung}
\affiliation{Department of Physics, The University of Texas at Austin, Austin, Texas 78712, USA}
\author{Gregory A. Fiete}
\affiliation{Department of Physics, The University of Texas at Austin, Austin, Texas 78712, USA}
\date{\today}

%%%%%%%%%%%%%%%%%%%%%%%%%%%%%%%%%%%%%%%%%%%%%%%%%%%%%%%%%%%%%%%
\begin{abstract}
Using a combined perturbative and self-consistent mean-field approach that we directly compare with quantum Monte Carlo calculations, we study the effects of short-ranged interactions on the $Z_2$ topological insulator phase, also known as the quantum spin Hall phase, in two generalized versions of the Kane-Mele model at half-filling on the honeycomb lattice.  For interactions weaker than the critical value for magnetic instability, we find that the interactions can stabilize the quantum spin Hall phase against third neighbor hoppings, which preserve $C_3$ lattice rotation symmetry, but destabilize it for a dimerization that explicitly breaks the $C_3$ symmetry.  Consistent with quantum Monte Carlo calculations, we show the phase boundary shifts are linearly proportional to the square of the interaction strength, but with opposite sign--a result that cannot be reproduced with a perturbative treatment that does not also include a self-consistent treatment of the perturbed Hamiltonian.  Our results emphasize that short-range interactions can have subtle effects on the stability of topological phases, and may need to be treated by methods analogous to those we use here.
\end{abstract}
\maketitle

%%%%%%%%%%%%%%%%%%%%%%%%%%%%%%%%%%%%%%%%%%%%%%%%%%%%%%%%%%%%%%%%%%%%%%%%%
\textit{Introduction}--Among the exotic states of matter discovered in recent years, topological insulators (TI) are especially noteworthy for their novelty and potential technological applications~\cite{fu2007,moore2007,moore2010,roy2009,hasan2010,qi2011}.  Shortly after the prediction~\cite{Bernevig2006}, the first experimental realization of a time-reversal symmetry (TRS) protected quantum spin Hall system was reported in HgTe/(Hg,Cd)Te quantum wells~\cite{Konig2007,Roth:sci09}.  In all the accepted experimental examples of TI to date, the presence of the topological state and most of its properties can be well understood within a noninteracting model.  However, it is generally believed that interactions can lead to qualitatively new topological phenomena in both two~\cite{Young:prb08,Neupert:prb11,Qi11,Levin:prl09,Maciejko:prb13,Ruegg:prl12} and three dimensions~\cite{Kargarian:prl13,Maciejko:prl14,Pesin:np10,Kargarian:prb11,Maciejko:prl10,Wan:prb11,Go:prl12}.  In two-dimensions, the Kane-Mele (KM) model~\cite{kane2005a} has played an especially important role in the study of  $Z_2$ TI (also known as quantum spin Hall (QSH) insulators). The KM model consists of two time-reversed copies of the Haldane model~\cite{haldane1988} on the two-dimensional (2D) honeycomb lattice, with real first-neighbor hopping and  imaginary second-neighbor hopping arising from spin-orbit coupling (SOC).  To study interactions, the KM model has been supplemented with an onsite Hubbard $U$-term--the Kane-Mele-Hubbard (KMH) model--and investigated extensively, particularly with quantum Monte Carlo (QMC) which is free of the fermion sign problem~\cite{rachel2010, yu2011,zheng2011,hohenadler2011,Budich:prb12,wuwei2012,hohenadler2012,Griset2012,lang2013}.  Its phase diagram is now well understood. 

Recently, several fermion sign-free extensions of the KMH model have been proposed and studied with QMC~\cite{Hung2013,hung2014} with goal of understanding short-ranged interaction effects on the hopping-parameter-driven $Z_2$ topological phase transitions at half-filling. In this work, we study two of them, given by Eq.~\eqref{eq:H_G} and Eq.~\eqref{eq:H_D}, supplemented by a Hubbard-$U$ term, 
\begin{equation}
\label{eq:H_U}
H_U=U \sum_j n_{j\uparrow} n_{j\downarrow},
\end{equation}
where $n_{j \sigma}$ is the number of electrons on site $i$ with spin $\sigma$. Both models  preserve discrete particle-hole symmetry (PHS). For interaction strengths below the regime of magnetic instabilities, the QMC results show that the interactions produce a shift in the location of the phase boundary (opposite directions for the two models)~\cite{hung2014}. In this Letter, we examine the two models using perturbation theory followed by a mean-field Hartree-Fock decoupling scheme (MFHF) and directly compare the results to QMC. We find that the sign of the shift and linear scaling with $(U/t)^2$ are accurately reproduced by the combination of perturbation theory and a self-consistent calculation, though they are not captured by either one independently. Our results emphasize that short-range interactions can have subtle effects on the stability of topological phases, and may need to be treated by methods analogous to those we use here when other approaches are not available or desirable.

\begin{figure}
\centering
   \subfigure{\includegraphics[width=1.6 in]{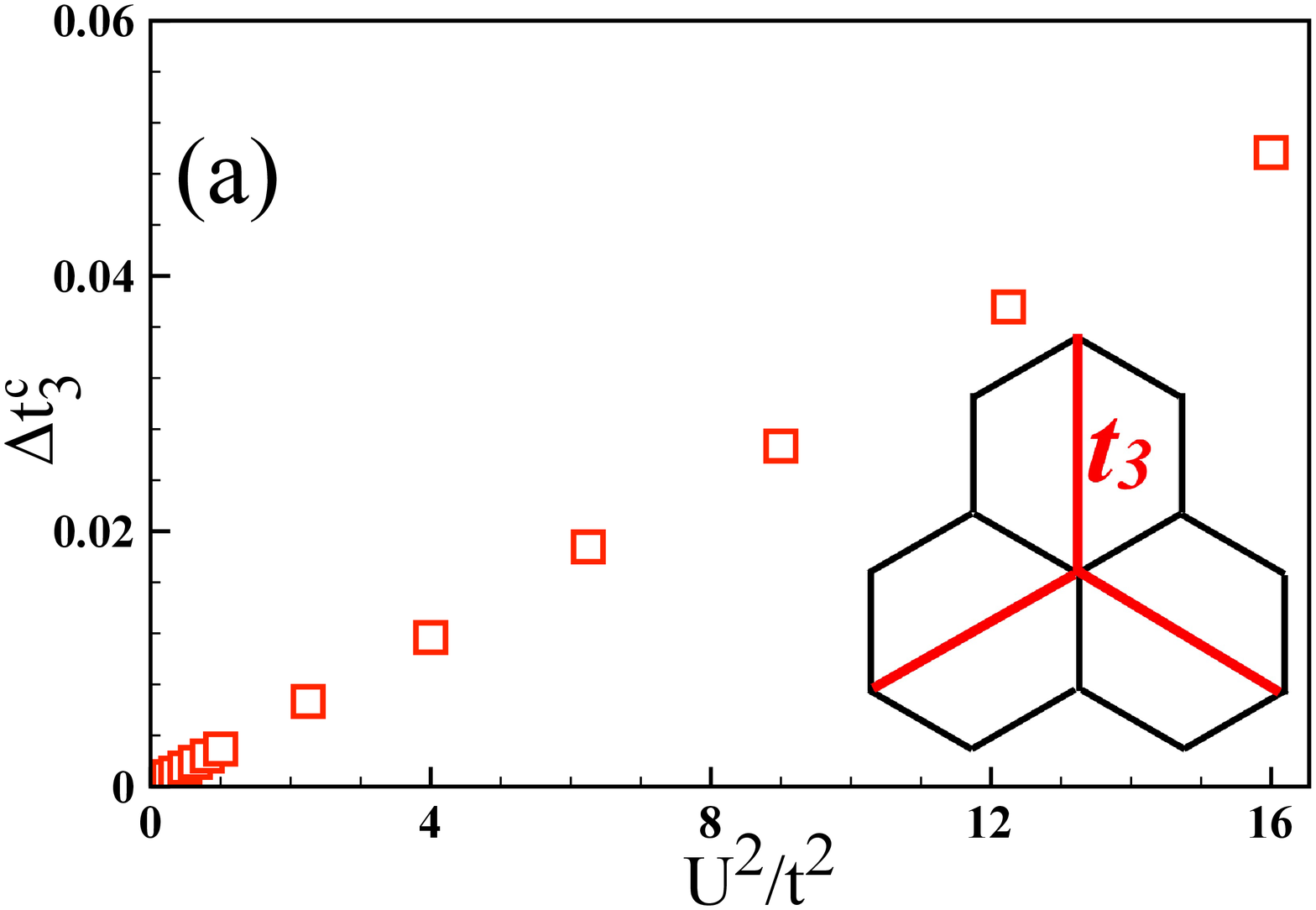}
      \label{Fig:GKM}}
   \subfigure{\includegraphics[width=1.6 in]{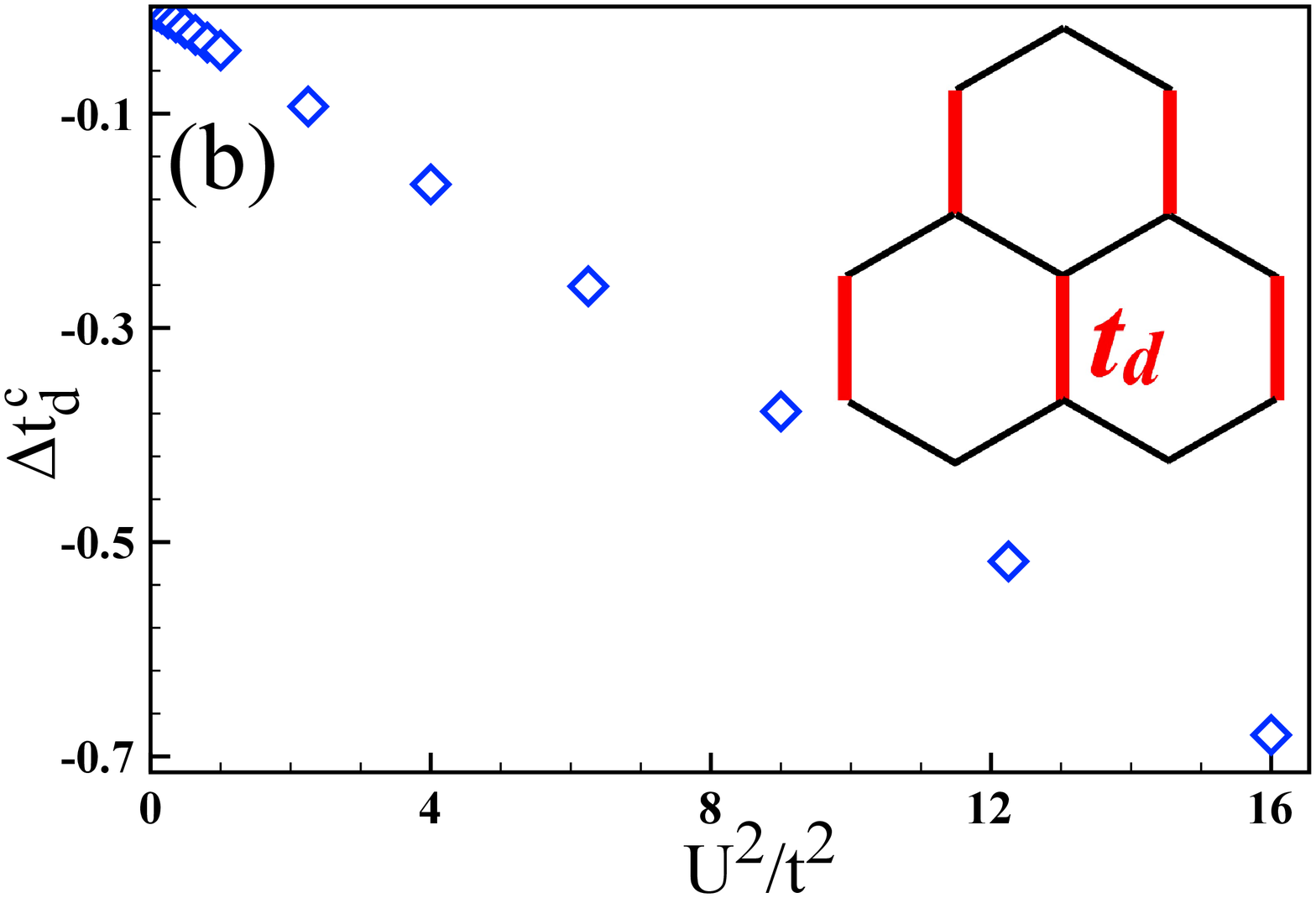} 
     \label{Fig:DKM}}
\caption{Self-consistent mean-field data for QSH boundary shift in the (a) GKM model and (b) DKM model within the perturbation theory plus mean-field picture. (a) The amount of the boundary shift, red open squares, is linearly proportional to $U^2/t^2$. The inset is the illustration of GKM model on the honeycomb lattice. The red lines represent the $t_3$ hoppings. (b) The open blue diamonds represent the data of the shift amount. The inset represent the DKM model on the honeycomb lattice with anisotropic hoppings that breaks $C_3$ rotation. The red lines represent the $t_d$ hoppings with $t_d > t$. A positive shift indicates the TI phase is stabilized; a negative shift it is destabilized.}
\label{Fig:MFHF}
\end{figure}

\textit{Variants of the KM model}--The first model we examine is the generalized Kane-Mele model (GKM)~\cite{Hung2013},  which includes real-valued third neighbor hoppings in addition to the original KM model, as illustrated in the inset in Fig.~\ref{Fig:GKM}. The second model is the dimerized Kane-Mele model (DKM)~\cite{lang2013}, which consists of anisotropic hoppings with hopping strength $t_d$ within a unit cell larger than those between different unit cells, inset in Fig.~\ref{Fig:DKM}. The GKM Hamiltonian, $H_{G}$, is
\begin{equation}
\label{eq:H_G}
H_{G}=  -\sum_{jk } \sum_{\sigma}  t_{jk} c^\dagger_{j \sigma} c_{k \sigma} + i \lambda_{so} \sum_{\la\la jk \ra\ra } \sum_{\sigma} \sigma c^\dagger_{j \sigma} \nu_{jk} c_{k\sigma},
\end{equation}
with $t_{jk} = t$ for $j, k \in \la j k \ra$, $t_{jk} = t_3$ for $j, k \in \la\la \la j k \ra \ra \ra$, and zero else, where $\la...\ra$, $\la\la...\ra\ra$, and $\la\la\la...\ra\ra\ra$ represent the nearest neighbors, the second neighbors, and the third neighbors. $\nu_{jk} = +1(-1)$ for (counter-)clockwise second-neighbor hopping and without lack of generality, we choose $t,\lambda_{so},  t_3 >0$.  The operator $c^\dagger_{i \sigma}$ ($c_{i\sigma}$) creates (annihilates) an electron on site $i$ with spin $\sigma$.  The DKM Hamiltonian, $H_D$, is
\begin{equation}
\label{eq:H_D}
H_{D}=  -\sum_{\la jk \ra} \sum_{\sigma} t_{jk} c^\dagger_{j \sigma} c_{k \sigma} + i \lambda_{so} \sum_{\la \la jk \ra \ra } \sum_{\sigma} \sigma c^\dagger_{j \sigma} \nu_{jk} c_{k\sigma},\\
\end{equation}
where $t_{ij} = t_d~(t)$ if the two sites $\la j k \ra$ belong to the same (different) unit cell(s), and we choose $t_d~(t)>0$.
\begin{figure}[t]
   \centering
   \includegraphics[width=1.6 in]{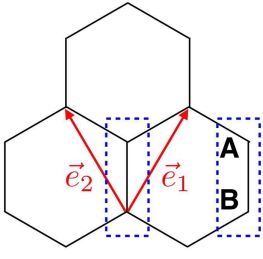}
   \caption{Schematic of the honeycomb lattice with two sublattices  labeled $A$ and $B$. The vectors $\vec{e}_{1/2} = (\pm 1/2, \sqrt{3}/2)$ connect the same sublattice in different unit cells.  The lattice constant is set to 1.}
  \label{Fig:honeycomb}
\end{figure}

From here forward we replace the site labeling $j$ with $j = \{{\bf r}, a\}$, where ${\bf r}$ runs over the Bravais lattice of unit cells of the honeycomb network and $a$ runs over the two sites ($A$ and $B$) in the unit cell shown in Fig.\ref{Fig:honeycomb}. The two different KM variants can be expressed in momentum space as $H_{\lambda} = \sum_{{\bf k}\in {\bf B. Z.}} \Psi^\dagger_{\bf k}\cdot h_{\lambda} \cdot \Psi_{\bf k}$, where $\lambda=$ G and D stand for GKM and DKM.  One has $h_\lambda = \mathbbm{M}_{\lambda}({\bf k})\otimes \mathbbm{1}_{2\times 2} +2\lambda_{so} g({\bf k}) \tau_z \otimes \sigma_z$, with 
\begin{eqnarray}
&&  \mathbbm{M}_{G}({\bf k}) = \begin{pmatrix}
0 & -t f({\bf k}) - t_3 f_3({\bf k}) \\
-t f^*({\bf k}) - t_3 f_3^*({\bf k}) & 0
\end{pmatrix},\\
&& \mathbbm{M}_{D}({\bf k}) = \begin{pmatrix}
0 & -t_d - t f_d({\bf k}) \\
-t_d -t f_d^*({\bf k}) & 0
\end{pmatrix},
\end{eqnarray}
where $\sigma_z$ and $\tau_z$ are the Pauli matrices for spin and sublattice degrees of freedom, and $ \Psi^{T}_{\bf k} \equiv \begin{pmatrix} \Psi^{\uparrow T}_{\bf k} & \Psi^{\downarrow T}_{\bf k} \end{pmatrix} = \begin{pmatrix}
c_{{\bf k} \uparrow}(A) & c_{{\bf k} \uparrow} (B) & c_{{\bf k} \downarrow} (A) & c_{{\bf k}\downarrow} (B)
\end{pmatrix}$, $g({\bf k}) \equiv -\sin ({\bf k}\cdot \vec{e}_1) + \sin({\bf k}\cdot \vec{e}_2) + \sin [{\bf k} \cdot (\vec{e}_1 - \vec{e}_2)]$, $f({\bf k}) = 1 + e^{i {\bf k} \cdot \vec{e}_1} + e^{i {\bf k} \cdot \vec{e}_2}$, $f_3({\bf k}) = e^{i {\bf k}\cdot(\vec{e}_1 + \vec{e}_2)} + 2\cos[{\bf k}\cdot(\vec{e}_1 - \vec{e}_2)]$, and $f_d({\bf k}) = e^{i {\bf k}\cdot \vec{e}_1} + e^{i {\bf k} \cdot \vec{e}_2}$.

%%%%%%%%%%%%%%%%%%%%%%%%%%%%%%%%%%%%%%%%%%%%%%%%%%%%%%%
\textit{Low energy description}--In the noninteracting limit, the band gap closes at time-reversal-invariant momenta (TRIM), located at ${\bf M}_{1,2} \equiv (\pm \pi, \pi/\sqrt{3})$ and ${\bf M}_3\equiv (0, 2\pi/\sqrt{3})$ \cite{hung2014}. At the TRIM, the diagonal elements of the Hamiltonian matrices vanish, $g({\bf M}_{a=1,2,3}) = 0$, and the band gaps in these models are actually controlled by the off-diagonal elements. For the GKM, we find that the gaps close at all three independent TRIM while for the DKM the gap only closes at one, say ${\bf M}_3$. Since the TRS relates the spin $\sigma$ and $\bar{\sigma}$, $\mathcal{T} : \Psi^\sigma_{\bf k} \rightarrow \epsilon^{\sigma \bar{\sigma}} \Psi^{\bar{\sigma}}_{-{\bf k}}$ with $\sigma = \uparrow(\downarrow) = 1~(2)$, we can simply focus on one spin species of fermions to have a complete description of the physics. The low-energy descriptions at the gap-closing points for spin $\sigma$ in each model are \cite{hung2014}
\begin{equation}
\begin{array}{lr} 
\label{eq:H_low}
\mathcal{H}^\sigma_G = \Delta t_G \Psi^{\sigma\dagger}_{{\bf M}_a} \tau_x \Psi^\sigma_{{\bf M}_a};&~~~ \mathcal{H}^\sigma_D = \Delta t_D \Psi^{\sigma\dagger}_{{\bf M}_3} \tau_x \Psi^\sigma_{{\bf M}_3},
\end{array}
\end{equation}
where we introduce $\Delta t_G = t-3t_3$ and $\Delta t_D = 2t - t_d$. Hence the band gaps are 
\begin{equation}
\begin{array}{lr}
 \Delta \mathcal{E}_{G} = 2 \left| \Delta t_G\right|; &~~~  \Delta \mathcal{E}_{D} = 2 \left| \Delta t_D \right],
 \end{array}
\end{equation}
which vanish at $\Delta t_G, \Delta t_D = 0$ ($t^c_3 = 1/3t$ for GKM and $t^c_d  = 2t $ for DKM)  \cite{hung2014}.

%%%%%%%%%%%%%%%%%%%%%%%%%%%%%%%%%%%%%%%%%%%%%%%%%%%%%%%
\textit{$U/t$ expansion and mean-field decouplings}--According to the low-energy descriptions in Eq.\eqref{eq:H_low},  the gaps vanish at the TRIM, unlike the usual Kane-Mele model, and are controlled by the off-diagonal elements describing the hopping between different sublattices.  In order to describe a possible shift of the topological phase boundary due to the presence of the short-range Hubbard interaction, a mechanism that can renormalize the off-diagonal elements (hopping amplitudes) of the Hamiltonian matrices is needed. A straightforward expansion in $U/t$ up to first order, $O(U/t)$, using the MFHF only gives an overall density correction which renormalizes the chemical potential without renormalizing the bare hopping amplitudes.  In order to capture the essential physics of the topological phase boundary shift, we perform the expansion in $U/t$ up to second order and then apply MFHF.  We find the $O(U^2/t^2)$ terms indeed give corrections to the off-diagonal terms consisting of the hopping correlations which can renormalize the bare hopping amplitudes leading to a shift of the $Z_2$ topological phase transition.

Performing the expansion in $U/t$ up to second order, we obtain the contributions to the bare Hamiltonians as $\delta H = \delta \mathcal{H}_1 + \delta\mathcal{H}_2$, where the $\delta \mathcal{H}_{1(2)}$ represent the first (second) order corrections. $\delta \mathcal{H}_1$ under straightforward mean-field HF decouplings gives
\begin{eqnarray}
\delta\mathcal{H}_1   \simeq  \frac{U}{2} \sum_{{\bf r},a=A,B} \bigg{[} \left\la n( {\bf r},a) \right\ra n({\bf r},a) + \left\la s_z ({\bf r},a) \right\ra s_z ({\bf r},a) \bigg{]},\nonumber \\ ~~ \label{Eq: Hubbard_1st_order}
 \end{eqnarray}
with $n\equiv n_\uparrow + n_\downarrow$, and $s_z \equiv n_\uparrow - n_\downarrow$.
We have explicitly neglected the constant $\la n_j \ra^2$ appearing in the MFHF since it only shifts the total energy. The terms $\la c^\dagger_\sigma c_{\bar{\sigma}}\ra $ also vanish since they do not conserve $S^z$. Since there is no local magnetic field at each site, the local magnetization is zero, which means the second term in Eq.~(\ref{Eq: Hubbard_1st_order}) vanishes. The on-site interaction within the MFHF picture simply renormalizes the diagonal elements of the Hamiltonian matrices. As QMC does not find a charge density wave state, we preserve the translational symmetry and set $\la n({\bf r},a) \ra= \la n (a)\ra \equiv \la n_a \ra$. In momentum space,  $\delta \mathcal{H}_1 ({\bf k}) = \sum_{{\bf k} \in B. Z.} \Psi^\dagger_{\bf k} h_1({\bf k}) \Psi_{\bf k}$, with 
\begin{eqnarray}
h_1 ({\bf k}) = \frac{U}{2}\begin{pmatrix}
 \la n_A \ra \\
 0 &  \la n_B \ra
 \end{pmatrix}
 \otimes \mathbbm{1}_{2\times2} = \frac{U\la n \ra}{2}  \mathbbm{1}_{4 \times 4},
 \end{eqnarray}
 where we explicitly used the fact that $\la n_A \ra = \la n_B \ra = \la n \ra$ above.
 
The second-order correction $\delta\mathcal{H}_2$ consists of two terms, $\delta\mathcal{H}_2 = \delta H^{(1)}_2 + \delta H^{(2)}_2$, with $\delta H^{(1)}_2 = -(U^2/2) \sum_{{\bf r}, {\bf r}', a}n_{\uparrow}({\bf r},a) n_{\uparrow}({\bf r}', a) n_{\downarrow}({\bf r},a)n_{\downarrow}({\bf r}',a),$ and $\delta H^{(2)}_2 = -U^2 \sum_{{\bf r}, {\bf r}'} n_\uparrow ({\bf r}, A) n_\uparrow ({\bf r}',B) n_\downarrow ( {\bf r}, A) n_\downarrow ({\bf r}',B)$. For simplicity in performing MFHF, we assume ${\bf r'} = {\bf r} + \vec{E}_{\mu}$, where $\vec{E}_\mu$ runs over the Bravais lattice of the unit cell that is connected to ${\bf r}$.  Then,
\begin{widetext}
\begin{eqnarray}
\delta H^{(1)}_2 =  && \frac{U^2}{2}  \sum_{{\bf k}, \vec{E}_\mu,\sigma, a}  \bigg{[} \bigg{(} \la n_\sigma(a)\ra \left| \chi_{\ell \bar{\sigma}}(\vec{E}_\mu, a ) \right|^2 - e^{-i {\bf k}\cdot \vec{E}_\mu} \chi_{\ell\sigma}(\vec{E}_\mu, a)  \left| \chi_{\ell \bar{\sigma}} (\vec{E}_\mu , a ) \right|^2 \bigg{)}  c^\dagger_{{\bf k} \sigma} (a) c_{{\bf k} \sigma}(a ) + \Hc\bigg{]},\label{Eq:H2-1}
\end{eqnarray}
\end{widetext}
where we define the $\ell$-neighbor hopping correlation $[\chi_{\ell \sigma}(\vec{E}_\mu,a)]^* \equiv \la c^\dagger_\sigma ({\bf r} + \vec{E}_\mu, a) c_\sigma ( {\bf r},a)\ra$, with $\ell$ being the number of sites covered by $\vec{E}_\mu$. For correlations between the same sublattices, $\ell$ is always even. For the GKM and DKM models, we restrict $\ell=2$ for second neighbor hopping (SOC) renormalization and $\vec{E}_\mu = \{\vec{e}_1,~\vec{e}_2,~\vec{e}_3 \equiv \vec{e}_1 - \vec{e}_2\}$, which is enough to capture the essential physics of the QMC results.  Under MFHF, $\delta H^{(1)}_2$ only renormalizes the diagonal terms of the Hamiltonian matrices.

For the MFHF of the $\delta H^{(2)}_2$, we introduce $({\bf r}',a) = ({\bf r}+ \vec{E}_\nu,a)$, with $\vec{E}_\nu$ being the vectors connected to ${\bf r}$. Note that $\vec{E}_\nu$ contain $\vec{e}_0 \equiv {\bf 0}$, which means the two sites are in the same unit cell. We obtain
\begin{eqnarray}
\nonumber && \delta H^{(2)}_2 =U^2 \sum_{{\bf k}, \vec{E}_\nu, \sigma} \bigg{[}  e^{-i {\bf k} \cdot \vec{E}_\nu}  \chi_{m \sigma} (\vec{E}_\nu, AB) \left| \chi_{m \bar{\sigma}}(\vec{E}_\nu,AB) \right|
^2\\
&& \hspace{4.7 cm} c^\dagger_{{\bf k} \sigma} (B) c_{{\bf k} \sigma} (A) + \Hc \bigg{]},\label{Eq:H2-2}
\end{eqnarray}
where $[\chi_{m \sigma} (\vec{E}_\nu, AB)]^* \equiv \la c^\dagger_\sigma ({\bf r} + \vec{E}_\nu , B) c_\sigma ({\bf r},A)\ra$, with $m$ being the number of sites covered by $\vec{E}_\nu$. Since $\vec{E}_\nu$ connects two different sublattices, $m$ is always odd. For simplicity, we restrict $m=1,~3$ for the GKM to capture the renormalizations of the first and third neighbor hoppings and $m=1$ for the DKM. For more efficient numerical calculations, we can utilize symmetries [$C_2$, Inversion + complex conjugation ($\mathcal{I}^*$), TRS, PHS for both GKM and DKM while there is an additional $C_3$ for GKM] to reduce the number of variables in each model.

\textit{(1) GKM model:} For the hoppings between different sublattices, we choose $\vec{E}_\nu = \{ \vec{e}_0 ,~\vec{e}_1, ~\vec{e}_2\}$ for $m=1$ and $\vec{E}_\nu = \{ \pm (\vec{e}_1 - \vec{e}_2),~\vec{e}_1 + \vec{e}_2\}$ for $m=3$. We can simplify Eqs. (\ref{Eq:H2-1})-(\ref{Eq:H2-2}) by identifying $\chi_{m \sigma}( \vec{E}_\nu ,AB) = \chi_{m \sigma}\equiv \chi_m$, $\chi_{2\sigma}(\vec{e}_1,a) = - \chi_{2\sigma}(\vec{e}_2,a) = - \chi_{2\sigma} (\vec{e}_1 - \vec{e}_2,a)$, and $\chi_{2\sigma} (\vec{e}_\mu ,a ) = \chi^*_{2\bar{\sigma}}(\vec{e}_\mu, a) = - \chi_{2\bar{\sigma}}(\vec{e}_\mu,a)$, where we use the fact that $\chi_{2\sigma}(\vec{e}_\mu,a) \in \mathbbm{I}$. For clarity, we introduce $\chi^{*}_{2\uparrow}(\vec{e}_1, a ) = i \chi_{2\uparrow}(a)$ and $\chi_{2\sigma}(A) = - \chi_{2\sigma}(B)\equiv \chi_2 \in \mathbbm{R}$. 

\textit{(2) DKM model:} For hopping between different sublattices we only need to consider the renormalizations of the first neighbor hoppings $t$ and $t_d$ with $m=1$.  Since the $C_3$ rotation is broken, the hopping amplitudes within the unit cell are no longer equivalent to those between different unit cells. Utilizing symmetry considerations, we can identify the hopping amplitudes within the same unit cell $\chi^\uparrow_1(\vec{e}_0,AB)=\chi^\downarrow_1(\vec{e}_0,AB) \equiv \chi^d_1$. For the hopping between different unit cells $\chi^\sigma_1(\vec{e}_1,AB) = \chi^\sigma_1(\vec{e}_2,AB)\equiv \chi_1$. For the second neighbor hopping, we have $\chi_{2\sigma} (\vec{e}_1, a) = - \chi_{2\sigma}(\vec{e}_2,a) \not = \chi_{2\sigma}(\vec{e}_3,a)$, $ \chi_{2\uparrow}(\vec{e}_{\mu=1,2,3}, a) = -\chi_{2\downarrow}(\vec{e}_\mu,a)$, and $\chi_{2\sigma}(\vec{e}_\mu, A) = - \chi_{2\sigma}(\vec{e}_\mu, B)$. For clarity, we define $\chi^*_{2 \uparrow}(\vec{e}_1, a) \equiv i \chi_{2\uparrow} (a)$, $\chi^*_{2\uparrow} (\vec{e}_3, a) \equiv i \chi^d_{2\uparrow}(a)$. We further introduce $\chi_{2\uparrow}(a) \equiv (-1)^{a+1} \chi_2 $ and $\chi^d_{2\uparrow}(a) \equiv (-1)^{a+1} \chi^d_2$, with $a=A~(B) = 1~(2)$.

\textit{Gap equations for the topological phase transition}--After utilizing symmetry arguments, we can self-consistently numerically solve for all parameters, $\chi$. For determining the shift of the phase transition location, we rely on the low-energy descriptions around the gap closing points, located at TRIM, and examine the gap equations below.

%%%%%%%%%%%%%%%%%%%%%%%%%%%%%%%%%%%%%%%%%%%%%%%%%%%%%%%%%%%%%%%
\textit{(1) Gap equation for GKM:}
\begin{eqnarray}\label{Eq:gap_GKM}
\Delta_{G} = t - 3 t_3 + U^2\bigg{(} \chi_1^3  - 3 \chi_3^3  \bigg{)}.
\end{eqnarray}
For the noninteracting critical point, $t_3 = 1/3 t$.  At weak-coupling, $U/t \ll 1$, we can approximate $\chi_1$ and $\chi_3$ to be the noninteracting values. We find that $\chi_1 \simeq 0.20705$ and $\chi_3 \simeq 0.03064$ and the $U^2$ correction is roughly $0.00879 U^2$. We conclude that at the weak-coupling limit, the topological phase is more stable against the third neighbor hoppings since we need larger $t_3$ to close the gap, consistent with QMC \cite{hung2014}.

\textit{(2) Gap equation for DKM:}
\begin{eqnarray}\label{Eq:gap_DKM}
\Delta_{D} = 2t - t_d - U^2 \bigg{[} (\chi^d_1 )^3 - 2 \chi_1 ^3 \bigg{]}.
\end{eqnarray}
Focusing on the critical point, $t_d = 2 t$, at the $U/t \ll 1$, we find that $\chi_1 \simeq 0.15770$ and $\chi^d_1 \simeq 0.36627$. The $U^2$ correction to the gap equation is $ -0.04129 U^2 < 0$. We conclude that at the weak-coupling regime the topological phase is more fragile to the dimerization, consistent with QMC \cite{hung2014}. 

 %%%%%%%%%%%%%%%%%%%%%%%%%%%%%%%%%%%%%%%%%%%%%%%%%%%%%%%%%%%%%%%%
\textit{Self-consistent numerical calculations}--In the self-consistent numerical calculations, the honeycomb lattice consists of $400 \times 400$ unit cells and we set $t=1$ and $\lambda_{so} = 0.4$. The results at finite $U/t$ for the GKM and DKM are shown in Figs.~\ref{Fig:GKM}-\ref{Fig:DKM}. The x-axis is the square of the interaction strength and the y-axis is the boundary shift amount, $\Delta t^c_3$ ($\Delta t^c_d$). We can see from Fig.~\ref{Fig:GKM}  that the on-site interaction stabilizes the QSH against the third neighbor hopping $t_3$ for GKM consistent with our previous weak-coupling picture, while Fig~\ref{Fig:DKM} shows that the interaction makes the QSH more fragile to the dimerization $t_d$ \cite{hung2014}. In addition, within MFHF, we find that the hopping amplitudes are almost independent of $U/t$ and, hence, the amount of boundary shift is linearly proportional to the $(U/t)^2$.

\begin{figure}
\centering
   \subfigure{\includegraphics[width=1.65 in]{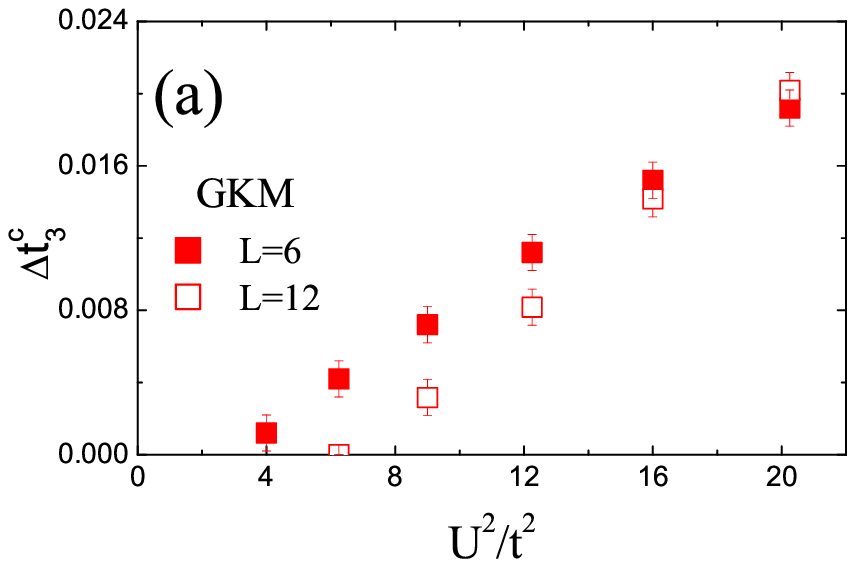}
      \label{Fig:GKM_QMC}}
   \subfigure{\includegraphics[width=1.6 in]{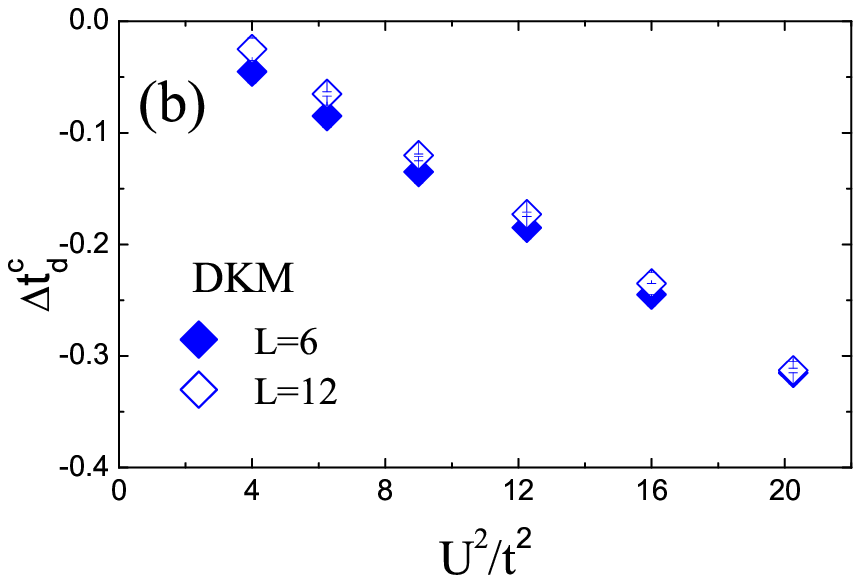} 
     \label{Fig:DKM_QMC}}
\caption{QMC data for QSH boundary shift in (a) GKM model and (b) DKM model. (a) The shift is positive, which means QSH is more stable against $t_3$ hoppings. More interestingly, the shift amount is linearly proportional to $(U/t)^2$, consistent with our mean-field picture. (b) The shift is negative and linearly proportional to $(U/t)^2$. The short-range interaction makes the QSH phase more destabilized by the dimerization $t_d$. Statistical errors are denoted by the error bars.}
\label{Fig:QMC}
\end{figure}

%%%%%%%%%%%%%%%%%%%%%%%%%%%%%%%%%%%%%%%%%%%%%%%%%%%%%%%%%%%%%%%%%%%%%%%%%%%%
\textit{Sign-free determinant projector QMC}--To further verify the MFHF, we perform sign-free QMC \cite{white1989,sorella1989,meng2010} for the GKM and DKM and plot them in Fig.~\ref{Fig:QMC}. The closed (open) red squares and closed (open) blue diamonds represent the boundary shifts ($\Delta t^c_3$ and $\Delta t^c_d$) obtained in $6\times 6$ ($12\times 12$) clusters, respectively. Due to the PHS, the Monte Carlo samplings in the KM variants are positive-definitive, and thus the results are numerically exact. Here we consider the discretized time step $\Delta \tau=0.05t$. The locations of the topological phase transition boundaries are characterized by the $Z_2$ index and spin Chern number, in terms of zero-frequency Green's functions~\cite{WangPRB, Wang_PRX, Wang2013, meng2013, hung2014}. In both models, the amounts and the signs of the boundary shift are linearly proportional to $(U/t)^2$ to high accuracy, consistent with the MFHF picture. Note that the linear relations to $(U/t)^2$ are universal and size-independent in the QMC results. 

\begin{figure}
\centering
 \includegraphics[width=1.6 in]{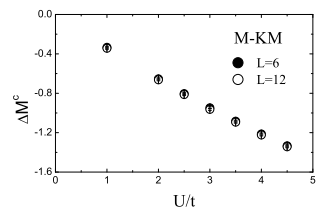}
\caption{QMC data for QSH boundary shift in the attractive Kane-Mele-Hubbard model in the presence of staggered potential $M$. The shift amount of the critical $M$, $\Delta M^c,$ is negative and linearly proportional to $U/t$. Note that we choose $\lambda_{so}=0.2t$, $U<0$ and PHS is explicitly broken due to the staggered potentials. QMC is only fermion-sign free in the attractive $U<0$ case. }
\label{Fig:QMC_staggerKM}
\end{figure}

%%%%%%%%%%%%%%%%%%%%%%%%%%%%%%%%%%%%%%%%%%%%%%%%%%%%%%%%%%%%%
\textit{Discussion}--The approach here is not unique to the KM-type models whose gaps close at TRI points. For an illustration, we consider a more conventional KM model in the presence of staggered potentials, $M_a=(-1)^{a+1}M$, which explicitly breaks PHS. Due to the staggered potentials, we know $\la n_B \ra \neq \la n_A \ra$. The gaps closing at the valleys ${\bf K}$ and ${\bf K}'$ with ${\bf K} = (4\pi/3,0)= - {\bf K}'$ in the Brillouin zone are controlled by the diagonal terms of the Hamiltonian matrix. For extracting the correct physics, we perform the $U/t$ expansion to first order and mean-field decouplings give an equation similar to Eq.~(\ref{Eq: Hubbard_1st_order}) with the diagonal elements replaced with $M_a + \la n_a\ra$. Following the same strategy above, we can define the gap function~\cite{Lai_staggerKM},
\begin{eqnarray}
\Delta_{s}({\bf K}) = 2M - 4 \lambda_{so}g({\bf K}) - \frac{U}{2}\bigg{(} \la n_B \ra - \la n_A \ra \bigg{)}.
\end{eqnarray}
For constant staggered potential $M$ and SOC $\lambda_{SO}$, the sign of the $U$ correction term is determined by the sign of $U$ (with $\la n_B \ra > \la n_A \ra$ assumed). For comparison with the QMC result, which can only work in the attractive interaction case, we choose $U = -|U|$. For $U<0$, the $U$ correction is positive. Since the QSH boundary is located at $\Delta({\bf K}) =0$, the critical staggered potential in the presence of interaction becomes smaller and decrease linearly as a function of $U/t$. The QMC data obtained in $6\times 6$ and $12\times 12$ attractive Kane-Mele-Hubbard model ($U=-|U|<0$ and $\lambda_{so}=0.2t$) in Fig.~\ref{Fig:QMC_staggerKM} explicitly confirms the MFHF's prediction. For other KM-type models, we believe our approach in this Letter, $U/t$ expansion $+$ MFHF $+$ low-energy gap equation $\Delta({\bf k})$, can essentially capture the interactions effects on the $Z_2$ topological phase transitions, and possibly in more general models as well.

%%%%%%%%%%%%%%%%%%%%%%%%%%%%%%%%%%%%%%%%%%%%%%%%%%%%%%%%%%
\textit{Conclusion}--We examined short-ranged interaction effects on two generalized versions of KM models, the GKM and DKM. We find that the interaction stabilizes the QSH in the GKM against the third neighbor $t_3$ hopping while makes the QSH more fragile to the dimerized hopping $t_d$ in the DKM. Within the mean-field Hartree-Fock picture, we conclude that the shift amount of the QSH boundary is proportional to $(U/t)^2$ and confirm this with exact QMC calculations. We believe this approach has a generally applicability when neither perturbative treatments nor mean-field treatments of the bare Hamiltonian exhibit stabilization/destabilization tendencies alone for the topological phase.

%%%%%%%%%%%%%%%%%%%%%%%%%%%%%%%%%%%%%%%%%%%%%%%%%%%%%%%%%%
\textit{Acknowledgments}--We are grateful to V. Chua, Z.-C. Gu, and L. Wang for collaborations on closely related projects, and for financial support from ARO Grant No. W911NF-09-1-0527 and NSF Grant No. DMR-0955778 (H.-H. Hung and G. A. Fiete) and NSF Grant No. DMR-1004545 (H.-H. Lai).
%%%%%%%%%%%%%%%%%%%%%%%%%%%%%%%%%%%%%%%%%%%%%%%%%%%%%%%%%%%%%
\bibliography{biblio4KM}
\end{document}